\documentclass[10pt,conference]{IEEEtran}

\usepackage{xcolor}

\usepackage{multirow} 
\usepackage{booktabs}

  \usepackage[pdftex]{graphicx}
\usepackage{algorithmic}
\usepackage{multirow} 
\usepackage{makecell} 
\IEEEoverridecommandlockouts
\makeatletter
\makeatletter
\def\ps@IEEEtitlepagestyle{
    \def\@oddfoot{} 
}
\let\old@ps@headings\ps@headings
\let\old@ps@IEEEtitlepagestyle\ps@IEEEtitlepagestyle

\def\confheader#1{%
  \def\ps@headings{%
    \old@ps@headings%
    \def\@oddhead{\strut\hfill#1\hfill\strut}%
    \def\@evenhead{\strut\hfill#1\hfill\strut}%
    \def\@oddfoot{} 
  }%
  \def\ps@IEEEtitlepagestyle{%
    \old@ps@IEEEtitlepagestyle%
    \def\@oddhead{\strut\hfill#1\hfill\strut}%
    \def\@evenhead{\strut\hfill#1\hfill\strut}%
    \def\@oddfoot{} 
  }%
  \ps@headings%
}
\makeatother
\makeatother
\confheader{} 
\begin{document}
%
%
\title{Resource Utilization of Differentiable Logic Gate Networks Deployed on FPGAs}
\author{\IEEEauthorblockN{Stephen Wormald, Gilon Kravatsky, Damon Woodard, Domenic Forte}
\IEEEauthorblockA{Florida Institute of National Security\\
University of Florida\\
Gainesville, Florida, USA\\
Email: stephen.wormald@ufl.edu}\\
}
%
%
%


\maketitle


\begin{abstract}

On-edge machine learning (ML) often strives to maximize the intelligence of small models while miniaturizing the circuit size and power needed to perform inference. Meeting these needs, differentiable Logic Gate Networks (LGN) have demonstrated nanosecond-scale prediction speeds while reducing the required resources as compares to traditional binary neural networks. Despite these benefits, the trade-offs between LGN parameters and resulting hardware synthesis characteristics are not well characterized. This paper therefore studies the tradeoffs between power, resource utilization, inference speed, and model accuracy when varying the depth and width of LGNs synthesized for Field Programmable Gate Arrays (FPGA). Results reveal that the final layer of an LGN is critical to minimize timing and resource usage (i.e. 28\% decrease), as this layer dictates the logic size of summing operations. Subject to timing and routing constraints, deeper and wider LGNs can be synthesized for FPGA when the final layer is narrow. Further tradeoffs are presented to help ML engineers select baseline LGN architectures for FPGAs with a set number of Look Up Tables (LUT).

\end{abstract}
\renewcommand\IEEEkeywordsname{Keywords}

%


\section{Introduction}
Edge machine learning (ML) increasingly demands models that are both computationally efficient and capable of high-speed inference under strict power and resource constraints \cite{kumar2024edge}. Traditional neural network architectures, while accurate, often require significant hardware resources and energy, limiting their deployment on small or embedded devices \cite{xu2023approx}. Differentiable Logic Gate Networks (LGNs) offer a promising alternative by combining the expressivity of neural networks with the resource efficiency of logic-based circuits \cite{petersen2022deep}. LGNs are neural networks trained to learn combinations of logic gates (i.e. AND, OR, XNOR gates, etc.) that easily map to hardware implementations, and have achieved nanosecond-scale prediction speeds \cite{petersen2022deep} and reduced resource requirements compared to traditional binary neural networks \cite{susskind2022weightless} while remaining explainable \cite{wormald2025explogic}, making them attractive for deployment on FPGAs and other low-power hardware accelerators.

\begin{figure}[h!]
    \centering
    \includegraphics[
        width=\linewidth,
        trim=9.75cm 9.5cm 9.25cm 9.5cm,
        clip
    ]{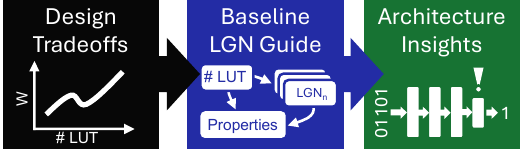}
    \caption{Overview of the paper, showing how design tradeoffs may be used to identify constraints for baseline LGNs and resulting architectural insights}
    \label{fig:overview}
\end{figure}

Despite these advantages, the relationships between LGN model size, inference latency, FPGA resource utilization, and power consumption remain poorly characterized. Engineers lack clear guidance for selecting network depth, width, and bit precision to satisfy specific design constraints. The work presented (see Figure \ref{fig:overview}) helps address this gap by systematically studying tradeoffs between LGN design variables and hardware metrics when targeting the Alveo-U200 \cite{xilinxU200}. Through a set of complementary studies evaluating accuracy, latency, resource usage, and power, this paper is the first to provide a practical methodology for mapping the number LUTs on an FPGA to baseline LGN architectures which may act as starting baselines when training LGNs for novel tasks. This paper is also the first to highlight the importance of final layers in LGNs when minimizing resource utilization on FPGAs as they connect to summation operations. The insights provided apply to a variety of edge computing applications, including low-power medical devices, autonomous vehicles, and embedded robotics, where efficient inference is critical.

\section{Methodology}
\label{sec:methods} 

Directed toward practical design insights, the methodology produces design tradeoffs between LGN architecture variations, model accuracy, circuit inference speed, resource utilization, and device power. A set of LGN models were trained to perform classification on common image datasets using the open source repository called DiffLogic \cite{petersen2022deep} before evaluating each model's accuracy (see Section \ref{sec:methods:lgn}). Each trained LGN was converted into hardware description language (HDL) code before performing synthesis (see Section \ref{sec:methods:synthesis}). In this study, all designs target the Alveo-U200 as it holds an XCU200 FPGA with a logic fabric containing 892,000 LUTs \cite{xilinxU200}, making it possible to test large LGN architectures. Resulting hardware kernels helped characterize the LGN performance tradeoffs, and are used to present a notional methodology for selecting baseline LGNs for novel applications (see Section \ref{sec:methods:tradeoffs}). 

\subsection{Design of LGN Architectures to Study Hardware Utilization}
\label{sec:methods:lgn} 

\begin{figure}[h!]
    \centering
    \includegraphics[
        width=\linewidth,
        trim=7.5cm 9cm 8cm 9cm,
        clip
    ]{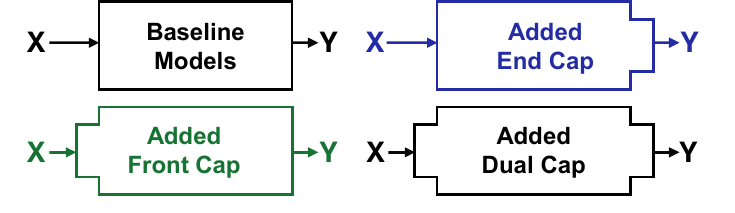}
    \caption{Summary of the model types trained using the DiffLogic library \cite{petersen2022deep}}
    \label{fig:model_types}
\end{figure}

LGNs were trained with key architectural variations to test the impact on synthesis and utilization. The experiment design considers that basic LGNs may be converted into HDL using three main blocks of logic which use separate LUT resources: (1) $LUT_{input}$ for input streaming, (2) $LUT_{logic}$ for combinatorial decision logic, and (3) $LUT_{sum}$ for performing classification based on the groupsum and argmax operations. Here, the total number of LUTs (or $LUT_{total}$) is roughly $LUT_{total} = LUT_{input} + LUT_{logic} + LUT_{sum} + \delta$, where $\delta$ consists of the logic needed for passing inputs and outputs to and from Double Data Rate memory (DDR) memory. Considering this division of resources, four LGN variations were designed per Figure \ref{fig:model_types}. Baseline models have a constant layer width ($L_W$) across assorted layer depths ($L_D$), where variations from this baseline either add ``end caps," ``front caps," or both (i.e. ``dual caps") as a strategy to reduce routing issues that emerged for larger models. This strategy is tested for larger values of $L_W$, as summarized in Table \ref{tab:lgn_params}. Here, model caps are appended to baseline models as additional layers. Front caps are set at constant widths $L_{front}$ while end caps are set as fraction of the baseline layer width (i.e. $f_{end}$, where $L_{end} = L_{W}*f_{end}$) to ensure layers are never less than half the width of prior layers, per recommendations from \cite{petersen2022deep}. The network parameters are varied per the ranges in Table \ref{tab:lgn_params}.      

\begin{table}[ht]
    \centering
    \caption{LGN design parameters with designated ranges}
    \label{tab:lgn_params}
    \begin{tabular}{|l|c|c|c|c|c|}
        \hline
        \textbf{Study} & \textbf{$L_{W}$} & \textbf{$L_D$} & \textbf{$L_{front}$} & \textbf{$\textbf{$f_{end}$}$}  \\
        \hline
        \textbf{Baseline} & [1000,64000] & [1,8] & -- & -- \\
        \textbf{Front Cap} & [8000,64000] & [3,6] & [2000, 4000] & -- \\
        \textbf{End Cap}   & [8000,64000] & [3,6] & -- & [0.5, 0.75] \\
        \textbf{Dual Cap}  & [8000,64000] & [3,6] & [2000, 4000] & [0.5, 0.75] \\
        \hline
    \end{tabular}
\end{table}

Models were trained on the MNIST \cite{lecun1998mnist}, FashionMNIST \cite{xiao2017fashionmnist}, and Cifar10 \cite{krizhevsky2009cifar10} datasets using a range of bit depth (i.e. $b$) of 1 through 8. Model training consistently used an $80/20$ split across training and testing data, using 5-fold cross validation and a batch size of 64. Models were trained using the Adam optimizer with early stopping with a patience of 10 epochs and a learning rate of 0.01. In total, 390 models passed both training and synthesis, consisting of 280 baseline model variations, and 110 variations using the cap variations (i.e. 26 front, 33 end, and 51 dual cap variations). The model to FPGA implementation conversion process is described in the following section. 

\subsection{Hardware Synthesis Targeting the Alveo U-200}
\label{sec:methods:synthesis} 

\begin{figure}[h!]
    \centering
    \includegraphics[
        width=\linewidth,
        trim=6cm 5.5cm 6cm 5.5cm,
        clip
    ]{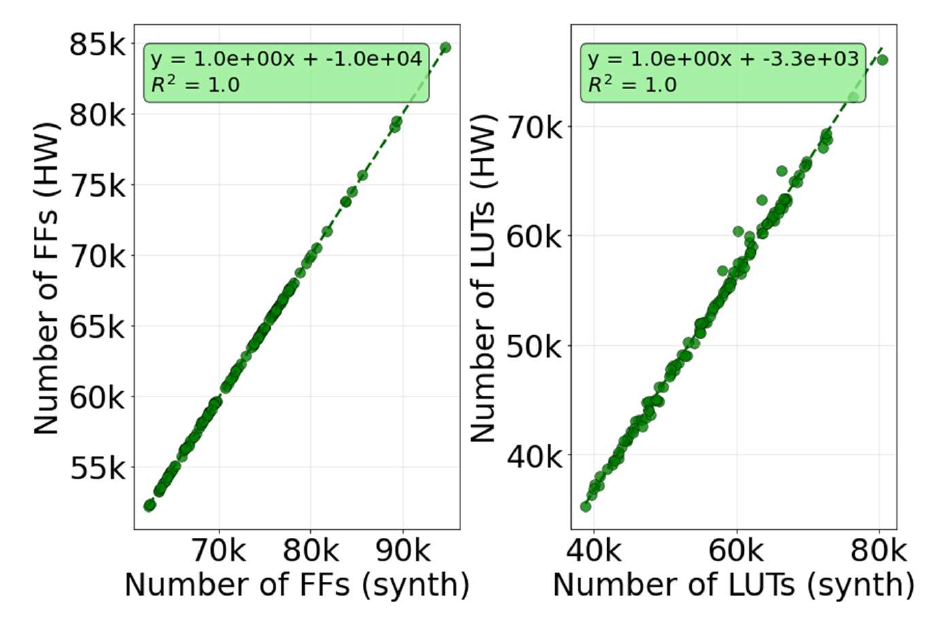}
    \caption{Comparing $FF$ and $LUT$ usage predicted in synthesis and realized in hardware, showing dependable synthesis predictions }
    \label{fig:hw_synth_comp}
\end{figure}

Each LGN model was converted into C++ and then into Verilog HDL using Xilinx Vitis HLS \cite{amd2025ug1399}. Note that larger models (i.e. when $L_{end} >\sim$100,000 gates) failed in pre-synthesis or experienced large synthesis times. When $L_{end}$ is large, the groupsum operations require additional LUT resources (i.e. $LUT_{sum}$). Hypothesizing that issues resulted from $LUT_{sum}$ resources inspired the design of the ``capped" experiments described in Section \ref{sec:methods:lgn}. The synthesized LUT usage ($LUT_{synth}$), Flip Flop (FF) usage ($FF_{synth}$), and number of compute cycles ($\# Cycles$) were calculated for each LGN to study design trade-offs. Hardware builds were performed for all jobs, but only 31.2\% of models finished this step due to failures during routing. Further inspection revealed that the The XCU200 FPGA inside the Alveo U200 accelerator uses stacked silicon interconnect (SSI) technology to combine three Super Logic Regions (SLRs) into a single logical device, effectively splitting the logic fabric across three die regions connected on an interposer. Routing between these dies produced congestion approximately at the point where the resources needed for a synthesized model could not all fit in a single SLR. As the number of LUTs and FFs on a hardware design ($LUT_{HW}$ and $FF_{HW}$, respectively) are linearly proportional to their synthesized counterparts (see Figure \ref{fig:hw_synth_comp}), $LUT_{synth}$ and $FF_{synth}$ are reported in Section \ref{sec:results}. The successful hardware designs inform the power tradeoffs.  

\subsection{Studying Tradeoffs Between LGN and Kernel Properties}
\label{sec:methods:tradeoffs} 

The synthesized LGNs were examined to identify tradeoffs between model and hardware properties. These tradeoffs consider: (1) \textit{Hardware design tradeoffs} which study the relationship between model size, performance, data input variables, and hardware utilization; (2) \textit{Kernel performance tradeoffs} which study the interplay between kernel power, latency, and hardware utilization; (3) \textit{LGN architecture tradeoffs} which reveal the correlation between LGN parameters and kernel properties; and (4) \textit{LGN performance tradeoffs}, which explore the relationship between resource utilization and model accuracy. These tradeoffs are combined into a notional methodology to help engineers identify application-specific baselines of LGN architectures for applications with known hardware and latency constraints. Results are provided in Section \ref{sec:results}.
\section{Results and Discussion}
\label{sec:results} 

The following sections present the LGN design tradeoffs alongside a methodology for designing LGN architectures when design constraints are known. At a high level, results show: (Section \ref{sec:results:accuracy}) $L_{W}$ and $L_{end}$ most strongly impact latency, likely due the impact on the $L_{sum}$ logic. (Section \ref{sec:results:power}) Power usage is minimal for the studied LGNs, and may generally be ignored for the range of LGNs studied except for highly efficient designs or large FPGA clusters. (Section \ref{sec:results:model}) LGN design freedom is high as few variables impact resource usage, aside from $L_{W}$, where $b$ and the total number of logic gates are useful for LGN design. (Section \ref{sec:results:caps}) End caps effectively reduce latency and resource utilization, and front caps are less impactful. Each result provides constraints when searching for LGN architectures as summarized in Section \ref{sec:results:method} and more generally discussed in Section \ref{sec:results:discussion}. 

\begin{figure}[h!]
    \centering
    \includegraphics[
        width=6.8cm,
        trim=0cm 0cm 0cm 0cm,
        clip
    ]{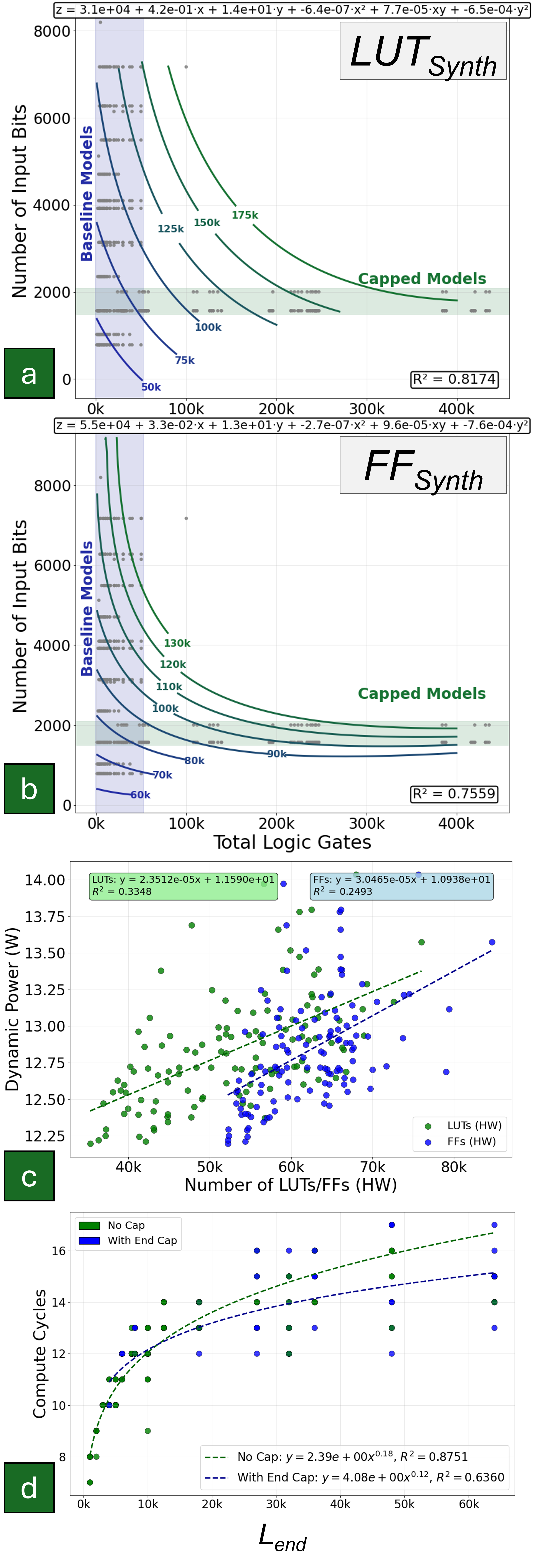}
    \caption{Synthesis design tradeoffs which may used to identify constraints when searching for baseline LGN architectures. Results show tradeoffs between the total number of LGN logic gates and the input data size for constant numbers of (a) LUTs and (b) FFs. Here, (c) the power increases with LUT and FF usage, and (d) the number of compute cycles mainly changes with $L_{end}$}
    \label{fig:main}
\end{figure}

\subsection{Hardware Design Tradeoffs: Accuracy, Latency, Resources}
\label{sec:results:accuracy} 

\begin{figure*}[h!]
    \centering
    \includegraphics[
        width=\linewidth,
        trim=0.cm 6.25cm 0.cm 6cm,
        clip
    ]{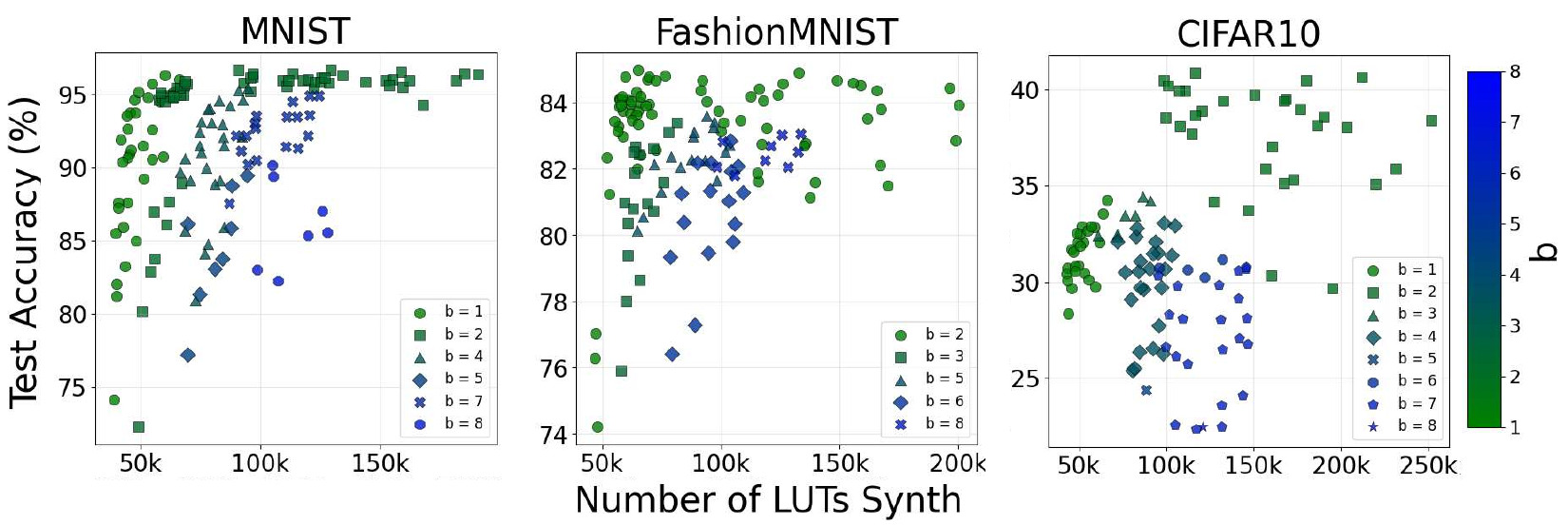}
    \caption{Accuracy change with respect to the model size, reported in the \#LUTs for synthesis. Here, and when b=1, MNIST and FashionMNIST have 784 bits per image, and CIFAR10 has 1024 bits per input}
    \label{fig:accuracy}
\end{figure*}

The objective of an LGN is generally to achieve high performance, or accuracy, while minimizing latency and resource usage for a given design. Here, accuracy generally increases with increased resource usage for a given input width (shown per the number of LUTs in Figure \ref{fig:accuracy}). However, models that spend resources processing larger inputs do not categorically perform better. For the studied models, LGNs trained with larger bit depths, or $b$, often performed worse while consuming additional resources. Deeper LGNs generally achieved better accuracies than shallow networks though $L_D$ has little-to-no impact on resource utilization and latency as the decision logic is implemented using combinatorial logic (see Table \ref{tab:correlation_matrix}), making $L_D$ valuable when increasing model accuracy, though results vary per dataset\footnote{Note that CIFAR10 represents an image dataset which is known to challenge feed forward LGNs \cite{petersen2022deep} which explains the relatively low accuracy}. In the context of this paper, accuracy is considered an objective rather than a constraint when designing baseline LGN architectures.  

The latency of a synthesized design is strongly correlated with the width of an LGN (Table \ref{tab:correlation_matrix}), often ranging from 7 to 17 cycles (Table \ref{fig:main}.d), which results from long data paths through a LGN. This finding is expected as $L_W$ determines the number of bits that are summed to classify an input, where large sums tend to rely on deep sequential logic. Pipelining can improve data throughput where latency is a less important. When latency acts as a design constraint, end caps provide a solution. Figure \ref{fig:main}.d shows how $\# Cycles$ increases logarithmically with $L_{end}$. End caps limit the size of the last layer, effectively moving the latency profile of an LGN to the left along this curve (e.g. by $\sim$2-3 clock cycles when using $f_{end} = 0.5$, per Table \ref{tab:combined_cap_effects}). Latency limitations may be used to identify constraints on $L_{end}$ per Figure \ref{fig:main}.d. 

\subsection{Kernel Performance Tradeoffs: Device Power Results}
\label{sec:results:power} 

The difference between the maximum and minimum power usage in LGNs examined is less than 2 Watts (Figure \ref{fig:main}.c). As such, the impact of potential power optimizations is minimal and may be less important as a constraint when designing baseline LGN architectures. Power optimization may be important for edge or low-power devices as well as large scale deployment in data centers. Though in these scenarios, other factors such as cost or overall power consumption may make FPGA implementations of LGNs less attractive.

\subsection{LGN Architecture Tradeoffs: Impact of LGN Variables}
\label{sec:results:model} 
The LGN architectural tradeoffs verify that $L_{end}$ is important to minimizing resource usage, which provides freedom in designing LGN architectures that maximize LGN-intelligence while minimizing FPGA resources. For example, Table \ref{tab:correlation_matrix} shows that $b$ and $L_D$ hold a weak correlation with $LUT_{Synth}$ when compared to $L_{end}$-associated variables (i.e. $L_{end}$, $L_{W}$, and $\#Gates$). Here, $L_{W}$ and $\#Gates$ impact resource usage because they either impact or are derived from $L_{end}$. This observation indicates that other variables, like $b$, $L_D$, or LGN wiring (i.e. how to connect two layers when routing information between LGN layers), need not constrain the search for LGNs targeting a given FPGA with limited LUTs. 


\begin{table}[h]
    \centering
    \caption{Correlation Matrix for Design Parameters Across both Baseline and End Cap Experiments}
    \label{tab:correlation_matrix}
    \begin{tabular}{|c|c|c|c|c|c|c|}
        \hline
        Metric & b & \#bits & $L_D$ & $L_W$ & $L_{end}$  & \#Gates\\
        \hline
        \textbf{$LUT_{Synth}$} & 0.38 & 0.28 & 0.36 & 0.35 & 0.80 & 0.72 \\
        \textbf{$FF_{Synth}$} & 0.76 & 0.64 & 0.12  & 0.30 & 0.33 & 0.28\\
        \textbf{$\#Cycles$} & -0.14 & -0.14 & 0.24  & 0.74 & 0.76 & 0.55\\
        \hline
    \end{tabular}
\end{table}

Figure \ref{fig:main}.a and Figure \ref{fig:main}.b seek to leverage this design freedom by providing constraints that help in searching for LGN architectures when limited by FPGAs with a set number of LUTs and FFs. Here, tradeoffs between the $\# Gates$ and $\# bits$ are plotted for constant values of $LUT_{Synth}$ and $FF_{Synth}$. While other variables correlate more strongly to resource usage, the tradeoff between $\#bits$ and $\# Gates$ helps engineers specify input properties (e.g. to help determine data collection and bandwidth properties) and set design budgets for the number of logic gates in certain LGN submodules. 

\subsection{Results of Model Caps on Accuracy and Resource Usage}
\label{sec:results:caps} 

While the tradeoffs between $\#bits$ and $\# Gates$ constrain the space of models compliant with a given FPGA, further constraints may may help optimize the performance of individual designs. Comparing the baseline models against front, end, and dual capped models helps explore such optimizations. 

End caps reduced latency and resource utilization more effectively than front caps (see Table \ref{tab:combined_cap_effects}). Specifically, models with end caps and without front caps reduced LUTs and FFs usage by $\sim$28$\%$ and $\sim$10$\%$, respectively. These models also reduced latency by up to 4 cycles when compared against baselines (or $\mu = 1.4 \pm 1.3$ (see Table \ref{tab:combined_cap_effects})). Note that front and end caps suffer a nominal accuracy penalty, generally in the range of 0-2$\%$ (see Table \ref{tab:combined_cap_effects}). Given the magnitude of improvement, LGN optimization routines may employ soft constraints which seek to minimize $L_{end}$ when a small reduction in accuracy is acceptable. Viewed alongside the prior constraints, these tradeoffs help in selecting LGN baselines for novel applications.  

\begin{table}[t]
    \centering
    \caption{Mean Change from Baselines ($Metric\pm$ Std. Dev.) Across All Capped Models (e.g. Front, End, Dual)}
    \label{tab:combined_cap_effects}
    \renewcommand{\arraystretch}{1.3}
    \scriptsize
    \begin{tabular}{|c|c|c|c|c|}
        \hline
        \textbf{Metric} & \textbf{$f_{end}$} & \textbf{$L_{front} = -$} & \textbf{$L_{front}=4k$} & \textbf{$L_{front}=2k$} \\
        \hline

        \multirow{3}{*}{\makecell{$LUT_{S.}$ \\ $(\%\Delta)$}}
        & --   & $0.0 \pm 0.0\%$ & $-6.0 \pm 14.7\%$ & $1.1 \pm 15.0\%$ \\
        & 0.75 & $-8.5 \pm 4.7\%$ & $-20.1 \pm 8.3\%$ & $-11.5 \pm 3.3\%$ \\
        & 0.5  & $-28.1 \pm 14.4\%$ & $-11.0 \pm 6.5\%$ & $-23.2 \pm 7.7\%$ \\
        \hline

        \multirow{3}{*}{\makecell{$FF_{S.}$ \\ $(\%\Delta)$}}
        & --   & $0.0 \pm 0.0\%$ & $-1.3 \pm 5.6\%$ & $-0.5 \pm 6.7\%$ \\
        & 0.75 & $-2.0 \pm 3.7\%$ & $-7.5 \pm 4.4\%$ & $-2.5 \pm 3.6\%$ \\
        & 0.5  & $-10.2 \pm 6.3\%$ & $-4.2 \pm 6.7\%$ & $-10.4 \pm 6.2\%$ \\
        \hline

        \multirow{3}{*}{\makecell{\#Cycles \\ $(\Delta)$}}
        & --   & $0.0 \pm 0.0$ & $0.0 \pm 1.2$ & $0.5 \pm 0.9$ \\
        & 0.75 & $0.1 \pm 0.8$ & $-1.2 \pm 0.9$ & $-0.1 \pm 1.0$ \\
        & 0.5  & $-1.4 \pm 1.3$ & $-0.2 \pm 1.3$ & $-1.5 \pm 0.9$ \\
        \hline

        \multirow{3}{*}{\makecell{Accuracy \\ $(\Delta)$}}
        & --    & $0.0 \pm 0.0\%$ & $-0.2 \pm 1.3\%$ & $-0.3 \pm 1.5\%$ \\
        & 0.75 & $-0.3 \pm 1.7\%$ & $0.3 \pm 1.3\%$ & $-0.5 \pm 2.0\%$ \\
        & 0.5  & $-0.2 \pm 1.6\%$ & $-0.7 \pm 2.9\%$ & $-1.0 \pm 2.6\%$ \\
        \hline
    \end{tabular}
\end{table}

\subsection{Resulting Method to Select Baseline LGN Architectures}
\label{sec:results:method} 

\begin{figure}[h]
    \centering
    \includegraphics[
        width=\linewidth,
        trim=9cm 7.4cm 10.5cm 8cm,
        clip
    ]{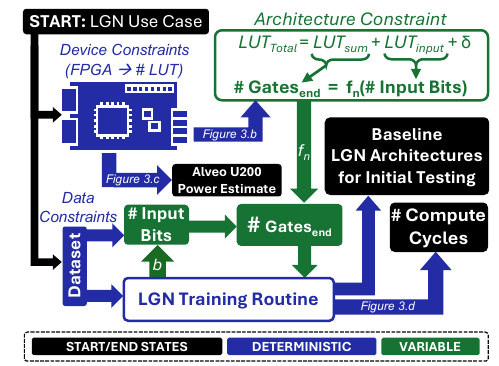}
    \caption{Flowchart showing how key results from Figure \ref{fig:main} may be used in a notional optimization routine when designing baseline LGN parameters}
    \label{fig:application}
\end{figure}

Synthesizing the design tradeoffs from the prior results sections produces a notional design methodology for identifying candidate LGN architectures based on a FPGAs available resources, design latency constraints, and target power requirements. This methodology integrates the results from each study as follows, and as summarized in Figure \ref{fig:application}: 

\begin{itemize}
    \item {\textit{Step 1 - Identify Edge Application and Data Constraints}}: For a given edge application, any data constraints should be identified (e.g. data size, bandwidth, etc.). 
    \item {\textit{Step 2 - Identify Available Resources}}: An engineer may determine the number of LUTs and FFs available on a candidate FPGA device. These values may be used to approximate the maximum power draw for the Alveo-U200 from Figure \ref{fig:main}.c.
    \item {\textit{Step 3 - Determine LGN Architecture Constraints}}: The selected FPGA may be used to determine either LUT or FF limitations and identify the corresponding tradeoffs between $\# bits$ and the $\# Gates$. If data constraints are known for a given application, a target number of logic gates may be derived for searching for LGN architectures. Else, the trend between $\# bits$ and the $\# Gates$ for a target LUT or FF may be derived from the corresponding equation and used as a constraint for architecture search. 
    \item {\textit{Step 4 - LGN Architecture Search}}: An optimization routine may be used to identify candidate LGN architectures for a given application using: (1) The tradeoff between $\# bits$ and the $\# Gates$; (2) A soft constraint to minimize $L_{end}$ from Section \ref{sec:results:caps}, or a hard constraint based on an objective latency defined as $\# Cycles$ from Figure \ref{fig:main}.d; (3) A design objective to maximize accuracy. Baseline architectures may be down-selected from the trace of feasible model designs explored by a given search algorithm.  
    \item {\textit{Step 5 - LGN Conversion and Optimization}}: Baselines may be synthesized to determine the true number of LUTs and FFs required for a given design, and used to update the target number of LUTs and design tradeoffs, per Step 3. This method may be repeated until achieving a LGN with suitable baseline characteristics.    
\end{itemize}

While variability is expected depending on the targeted FPGA, once setup the proposed methodology helps engineers quickly propose candidate LGN architectures and converge on application-specific designs.

\subsection{Discussion on Generalizability of Results}
\label{sec:results:discussion} 

While the prior methodology is designed using models trained for small-scale image classification and the accuracy results do not generalize beyond the studied datasets, the relationships found between model input, logic gates, and resource usage remain dataset agnostic. Several further limitations are of note to aid the use of results and inspire further study:
\begin{itemize}
    \item {\textit{{On Placement, Routing, and LUT Mapping}}: Two different FPGAs given the same synthesized design will yield different resource utilization's. Different FPGAs have different resources, such as LUT3 vs LUT4, and will not have the same hardware implementation. Furthermore, it is possible that when implementing a synthesized design on a single FPGA multiple times, the placement and or routing algorithms may result in a different implementation. Additionally, there are many possible LUT mappings for a given net list. Given the large combinatorial networks required for LGNs and the different LUTs available, it is unlikely that an implementation on another FPGA will have the same LUT mapping. For these reasons, the specific number of LUTs and FFs for a given LGN would differ on alternate FPGAs.}
\end{itemize}

\begin{itemize}
     \item \textit{{Power Disparity for Alternate FPGAs}}: The power usage results studied are generated by Vivado's power tools using our LGNs implementation on the Alveo-U200. The results are expected to hold some error when compared against real-world power usage, which also varies per clock speed. Since different FPGAs run at different clock speeds, the power usage for a given LGN would likely vary from this study. Additionally, the LUTs, FFs, and other resources on the Alveo-U200 will likely not have the same power usage as on another FPGA.  
     \item \textit{{On Timing Closure:}} To implement a design, it must meet timing requirements of the targeted FPGA. The timing of a design is a function of the target clock speed, the speed of LUTs and FFs, as well as the speed of resource interconnects. Give these variables, LGNs implemented on alternate FPGA will require unique techniques to optimize timing.
     \item \textit{{Method of input:}} Many architectural decisions to be made when implementing LGNs on FPGAs. This study employs a software accelerator model where a host PC would place input data in the Alveo U200's DDR memory via PCI-E. This data was retrieved and passed as input to the datapath, where it flowed through the LGN and other necessary elements. The output of the datapath was then written to DDR, where it was read from by the host PC via PCI-E. Fetching data from and writing data to memory is resource intensive and may vary between LGN implementations. For example, cameras connected to a FPGA could stream inputs to the data path via a FIFO buffer, which may be less resource intensive. These architectural decisions can impact the performance and resource utilization of a given LGN. 
\end{itemize}

In light of these shortcomings, the presented work is useful for establishing baseline LGN architectures, though device specific optimizations should be expected. The findings in this study hold value in minimizing the work engineers need to perform when working with novel applications of LGNs in edge environments. 

\section{Conclusion}

This study systematically studies tradeoffs between LGN design variables and the properties of synthesized hardware metrics, culminating in a notional methodology to identify baseline LGN architectures for novel applications. This study is the first to investigate the impact of LGNs final layers on hardware metrics. The value is in clarifying the influence of design and architectural variables when deploying LGNs on FPGAs. Focusing on small scale image classification tasks and synthesized hardware designs, this study helps establish a foundation for more robust LGN design and deployment strategies. In this study: 

\begin{itemize}
    \item 390 LGN variations were trained to test the impact on synthesis and hardware utilization (see Section \ref{sec:results:accuracy}), revealing that minimizing the end layer of a network (or $L_{end}$) can decrease latency while reducing LUT usage by up to 28\%. These results indicate that larger LGNs may be placed on relatively small FPGAs by reducing $L_{end}$.  
    \item HW power usage was analyzed for 122 models to show the impact of increased resource utilization on increasing power demands. In this study power utilization is minimal ($\sim$12.5-13.5W, per Section \ref{sec:results:power}).
    \item Design tradeoffs are provided for targeted FPGAs with known numbers of LUTs and FFs, helping engineers determine objective model sizes (i.e. the total number of gates) based on application-specific constraints on data (like the number of input bits, per Section \ref{sec:results:model}). These results help engineers budget the number of logic gates which may be used in LGN submodules.
    \item The design tradeoffs found through this study are synthesized into a notional methodology for selecting baseline LGN architectures when targeting FPGAs with known resource limitaions for novel applications (Section \ref{sec:results:method}).
\end{itemize}

These findings may be extended to compare LGN implementations on FPGAs and microcontrollers, evaluate design rules for LGNs designed into application specific integrated circuit (ASIC), and determine the value of LGNs architectural variations which are specialized for certain applications.
\begin{itemize}
    \item  There is room to explore and compare hardware utilization across smaller FPGAs and microcontrollers which have fewer resources to develop upon optimization techniques which help in deploying LGNs on edge devices (extending work from Section \ref{sec:results:accuracy}-E). 
    \item The design tradeoffs in Section \ref{sec:results:accuracy}-D are specific to LGN implementations on FPGAs. Future work could investigate how LGNs may hold value in designing ASICs for applications where large LGNs are needed while maintaining super low power usage.
    \item The findings on LGN model variations in Section \ref{sec:results:caps} indicate that continued research into specialized LGN architectures may further optimize resource usage and latency for small devices. 
\end{itemize}

Following this work, ML engineers can implement and extend the design methodology outlined in Section \ref{sec:results:method} to specialized applications and devices which hold unique constraints. Streamlining this approach for niche tasks holds the potential to improve the intelligence density deployed on small devices, which holds value in numerous applications across the Internet-of-Things, micro-robotics including autonomous vehicles, and personalized devices in the medicine. 

\bibliographystyle{IEEEtran}
\bibliography{IEEEabrv,GOMACTech_LaTeX}

\end{document}